\documentclass[pdflatex,sn-mathphys-num]{sn-jnl}% Math and Physical Sciences Numbered Reference Style
\usepackage{graphicx}%
\usepackage{multirow}%
\usepackage{amsmath,amssymb,amsfonts}%
\usepackage{amsthm}%
\usepackage{mathrsfs}%
\usepackage[title]{appendix}%
\usepackage{xcolor}%
\usepackage{textcomp}%
\usepackage{manyfoot}%
\usepackage{booktabs}%
\usepackage{algorithm}%
\usepackage{algorithmicx}%
\usepackage{algpseudocode}%
\usepackage{listings}%
\usepackage{enumitem}     
\usepackage{amsmath}   % for \numberwithin
\usepackage{chngcntr}  % for \counterwithin

\usepackage{bm}% bold math
\usepackage{float}
\usepackage[utf8]{inputenc}
\usepackage[T1]{fontenc}
\usepackage{mathptmx}
\usepackage{etoolbox}
\usepackage{hyperref}
\usepackage{nameref}

\usepackage{graphicx}
\usepackage{subcaption}
\usepackage{epstopdf}

\theoremstyle{thmstyleone}%
\theoremstyle{thmstyletwo}%

\theoremstyle{thmstylethree}%

\begin{document}

\title[Article Title]{Experimental investigation of the water exit of the accelerating circular cylinder}


\author*[1,2]{\fnm{Intesaaf} \sur{Ashraf}}\email{intesaaf.ashraf@ucl.ac.uk}

\author[1]{\fnm{Stephane } \sur{Dorbolo}}\email{s.dorbolo@uliege.be}

\affil*[1]{\orgdiv{PtYX, Département de Physique}, \orgname{Université de Liège}, \orgaddress{\city{Liege}, \postcode{4000}, \country{Belgium}}}

\affil[2]{\orgdiv{Mechanical Engineering Department}, \orgname{University College London}, \orgaddress{\city{London}, \postcode{WC1E 7JE}, \country{United Kingdom}}}


\abstract{
We investigate the vortical dynamics generated during the forced exit of a circular cylinder through a quiescent free surface. While previous work on exit flows has largely emphasized cavity collapse and force generation, the formation and evolution of the starting vortex beneath the body have received much less attention. Using high-speed particle image velocimetry at up to 2000~Hz, we isolate and track the primary vortex shed during exit. The body motion is driven by constant vertical acceleration from rest, such that the cylinder kinematics satisfy $v_{\text{cyl}}^2 = 2\alpha\,\Delta y_{\text{cyl}}$, with $\Delta y_{\text{cyl}}$ the displacement since release. This controlled protocol allows systematic variation of $\alpha$ while keeping the maximum crossing velocity fixed. We characterize vortex trajectories, equivalent radius, peak vorticity, and circulation, and compare across accelerations. The results show that circulation growth collapses under the impulse-based scaling $\Gamma/\sqrt{\alpha\,(\Delta y_{\text{cyl}})^3}$, independent of the imposed acceleration. A consistent transition is observed as the vortex centroid approaches the free surface, where its evolution is modified by surface interaction. The trajectories remain predominantly vertical with only weak lateral drift, robust across all accelerations. Taken together, these findings establish a systematic framework for water-exit vortex dynamics under constant acceleration and highlight scaling features that parallel the classical starting-vortex problem in entry flows.

}




\maketitle

\section{Introduction}\label{sec1}

The exit of solid bodies from a liquid surface represents a systematic problem in unsteady fluid mechanics with broad relevance to naval architecture, aerospace recovery, ocean engineering, and dip-coating processes. These flows involve coupled dynamics of free-surface deformation, vortex formation, and hydrodynamic loading, making them central to understanding cross-media propulsion and surface-manipulation technologies.

While water-entry phenomena have been extensively studied \citep{challa2014, challa2010, mohtat2015, yang2012}, the complementary problem of water-exit has received less systematic attention \citep{Havelock1936, truscott2016, wu2017experimental, haohao2019numerical}. Interest in this regime has grown recently, motivated by the performance and control of submerged vehicles and bio-inspired systems. Experimental and numerical studies have sought to characterize forces, entrainment, and wake patterns during exit \citep{ashraf2024, ashraf2024effect, ashraf2024exit, takamure2025motion, huang2024numerical, zhou2024numerical}, but the governing mechanisms remain incompletely understood.

Early theoretical contributions include \citet{Havelock1936}, who analysed vertical cylinder motion in a quiescent fluid using potential flow. \citet{greenhow1983nonlinear} reported nonlinear surface elevations and chaotic wave behaviour during cylinder ascent, while \citet{telste1987} categorized exit scenarios based on exit velocity. Subsequent experiments expanded this foundation: \citet{greenhow1997water} studied two-dimensional flows, \citet{liju2001} analysed geometric effects on surge dynamics, and volume-of-fluid (VOF) modelling was applied to capture wave formation and air entrainment during non-vertical exits \citep{moshari2014, kleefsman2004improved, nair2018water}. In these studies, cavity dynamics and free-surface deformation were primary observables.

Work on spherical geometries highlighted additional mechanisms. \citet{chu2010} demonstrated that cavities form at cylinder ends, impacting force signatures during deceleration and interface crossing. \citet{haohao2019numerical} and \citet{ni2015} showed how Froude number controls detachment and vortex development. \citet{truscott2016} measured buoyant sphere wakes across different initial depths and Reynolds numbers, while \citet{wu2017experimental} reported columnar ejections during partial submergence. \citet{ashraf2024effect} found that dimpled spheres reduce drag and entrainment coefficients while preserving cross-over forces, revealing how surface topology modifies exit dynamics. More recently, \citet{takamure2025motion} introduced vented spheres with through-holes, which altered entrainment symmetry and induced rotation. \citet{zheng2025experimental} studied slender cylinders of varying head shape, aspect ratio, and stiffness, showing that conical heads improve cavity stability while flexibility influences trajectory control.

Complementary advances in numerical modelling have provided further insight into unsteady exit processes. Potential-flow and VOF simulations have captured surface deformation and cavity collapse \citep{greenhow1997water, moshari2014, kleefsman2004improved, nair2018water}. Detached Eddy Simulations (DES) and wave-tank experiments \citep{zhou2024numerical} identified cavity asymmetries and the role of vapor dynamics in force modulation. RANS–VOF simulations by \citet{huang2024numerical} demonstrated that wave-phase alignment can minimize pitch deviation during deep-sea mining vehicle (DSMV) exit. Large-eddy simulations of neutrally buoyant spheres \citep{huang2025dynamics} segmented the motion into submerged, partially submerged, and airborne stages, revealing phase-specific mechanisms of energy redistribution, including viscous damping, buoyancy loss, and ballistic motion. These models highlight the multiphase nature of exit flows and the sensitivity of cavity and wake structures to acceleration, shape, and boundary conditions.

Parallel progress has emerged in machine-learning-based prediction of water-exit behaviour. \citet{huang2025application} trained parallel neural networks (PNN) on CFD-derived datasets to predict DSMV trajectories under various wave phases, achieving high accuracy ($MSE < 0.01$, $R^2 > 0.99$) while reducing computational cost. \citet{li2025attention} proposed an Attention-UNet and Angle-CNN framework for simultaneous flow-field and six-degree-of-freedom orientation prediction, obtaining $SSIM > 0.90$ and $R^2 > 0.998$ for pitch-angle prediction. These studies underline the potential of data-driven approaches to accelerate predictions, but they remain fundamentally dependent on the accuracy of training data.

Despite this breadth of prior work, most exit studies have emphasized cavity morphology, hydrodynamic loads, or free-surface behaviour. Far less is known about the vortex dynamics beneath an accelerating body. In particular, the growth of circulation—a direct measure of hydrodynamic impulse transfer—has not been systematically measured during constant-acceleration exits. Existing experiments have mainly examined constant-speed withdrawal or wave-driven motions, leaving the impulsive, acceleration-controlled regime unexplored. Yet understanding circulation growth and scaling is essential, both as a systematic benchmark for free-surface CFD and as a foundation for predictive design rules in cross-media vehicles.

Here, we conduct time-resolved particle image velocimetry (2000~Hz) of a circular cylinder ($a=12.5$~mm) exiting quiescent water under constant vertical acceleration. The dynamics are parameterized by two acceleration-based non-dimensional numbers, 
\begin{equation}
Fr = \sqrt{\frac{\alpha}{g}}, 
\qquad 
Re = \frac{a^{3/2}\sqrt{\alpha}}{\nu},
\end{equation}
where $\alpha$ is the imposed acceleration, $a$ the cylinder radius, $g$ gravity, and $\nu$ the kinematic viscosity. The Reynolds number is defined using the characteristic velocity $U \sim \sqrt{\alpha a}$ that naturally arises from the impulse-based formulation 
(see Eq.~\ref{eq:GammaScaling}), giving $Re = Ua/\nu = a^{3/2}\sqrt{\alpha}/\nu$. This choice highlights the acceleration-controlled regime considered here, while reducing to the conventional form when $U$ is defined from a characteristic displacement. Accelerations $\alpha$ from $0.16$ to $4.0$~m\,s$^{-2}$ were imposed, corresponding to 
$Fr = 0.13$--$0.64$ and $Re = 5.6 \times 10^2$--$2.8 \times 10^3$.

From the measured spanwise vorticity fields, we extract vortex centroid, circulation, enstrophy, and equivalent radius throughout the ascent. We show that circulation collapses under an impulse-based scaling, whereas core-based normalisations succeed only in the earliest compact-core regime. This establishes impulse-based similarity as the governing framework for water-exit vortex dynamics, while providing a systematic dataset and quantitative benchmarks for unsteady free-surface simulations.

%This study aims to address these gaps by investigating the exit dynamics of a square cylinder pulled from water at a constant speed. During the process, interface deformation was measured to analyze the evolution of the water layer's thickness on the cylinder. Force measurements were synchronized with high-speed imaging. Two dimensionless parameters were employed: the Froude number, $Fr = \frac{U^2}{ga}$, where $U$ is the vertical velocity of the cylinder and $a$ is the cylinder's side length; and the Reynolds number, $Re = \frac{Ua}{\nu}$, where $\nu$ is the kinematic viscosity of water. The study focuses on the dynamics of interface crossing, considering factors such as object speed, shape, fluid viscosity, and surface tension. Emphasis is placed on the role of object geometry in determining exit behavior.

\section{Experimental methods}

\subsection{Experimental Set-up}
\begin{figure}
  \centering
  \includegraphics[width=\linewidth]{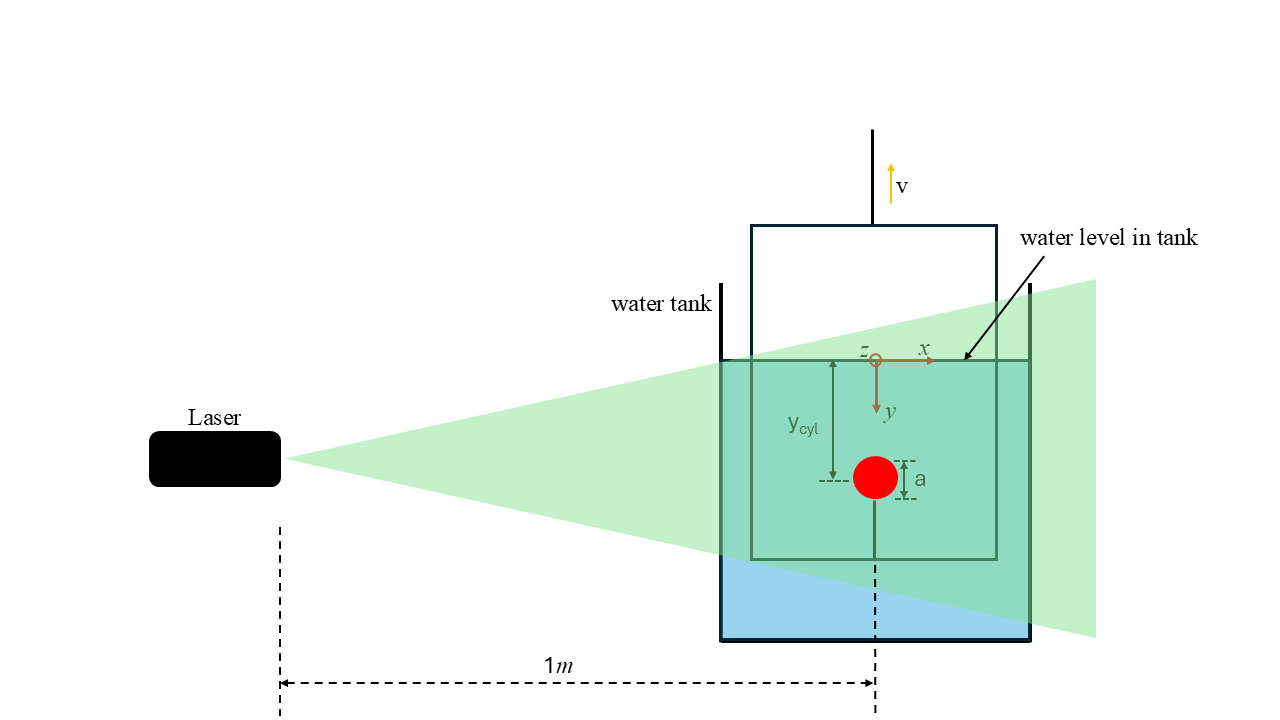}
  \caption{Schematic of the PIV setup. A vertical mid-plane was illuminated by a 532 nm laser sheet, with imaging at 2000 Hz.}
  \label{PIV_setup}
\end{figure}

The experiments were conducted in a transparent water tank of dimensions $78.5 \times 27.5 \times 72.5$ cm$^{3}$, with optical access from both the sides and front. Vertical translation of the test object was driven by a precision rack-and-pinion mechanism (figure~\ref{PIV_setup}). The object was a circular cylinder of radius $a=12.5$ mm and length $L=12a$, mounted horizontally on a rigid carbon fiber-rod frame to maintain alignment throughout the motion. The cylinder axis was parallel to the $x$-axis (along the tank length) and perpendicular to the direction of translation ($y$-axis).

Planar particle image velocimetry (PIV) was performed in the vertical mid-plane ($xy$-plane). A continuous-wave 532~nm, 4~W laser sheet illuminated neutrally buoyant tracer particles (diameter $\approx 20~\mu$m). Image sequences were acquired at 2000~fps with a field of view of $0.128 \times 0.205$~m$^{2}$, giving a spatial calibration of 6.075~pixels/mm. The laser sheet was positioned 1~m from the lifting mechanism to minimize optical distortion.  

Velocity fields were processed using PIVlab \citep{thielicke2014pivlab}, a MATLAB-based toolbox for digital PIV. Multi-pass cross-correlation with window deformation was applied, 
ending with $32 \times 32$~pixel interrogation regions at 50\% overlap. Spurious vectors were detected with a local median filter and replaced by bilinear interpolation. The estimated velocity uncertainty is $\approx 3\%$, consistent with the sub-pixel correlation accuracy reported for PIVlab.

The coordinate system was defined with origin at the undisturbed free surface along the pulling axis: $x$ along the tank length, $y$ vertically upward, and $z$ across the width. Spatial quantities are reported in units of $a$.  

The cylinder was impulsively accelerated from rest with prescribed constant accelerations up to $4.0$ m\,s$^{-2}$. Independent motion tracking verified the acceleration profile (see the Appendix~\ref{Cylinder kinematics} for details), and the PIV acquisition was synchronized with the steady acceleration phase. The initial depth was fixed at $10a$, ensuring that the acceleration profile was fully established before the interface was reached.  

\subsection{Vorticity and enstrophy}
From the velocity field $(u,v)$ ($u$ and $v$ are the speeds of the fluid along the $x$ and the $y$ directions, respectively), the spanwise vorticity was computed as
\begin{equation}
  \omega_z(x,y,t) = \frac{\partial v}{\partial x} - \frac{\partial u}{\partial y},
\end{equation}
with positive $\omega_z$ denoting counterclockwise rotation.  

Domain-integrated enstrophy was evaluated as
\begin{equation}
  E(t) = \iint_{\Omega} \omega_z^2(x,y,t)\,dA,
\end{equation}
where $\Omega$ denotes the interrogation window. Unlike circulation, enstrophy is strictly positive and highlights the formation and growth of vortical structures \citep{weiss1991dynamics, rivera2003energy, chen2003physical}.  

`NaN' velocity vectors were replaced by bilinear interpolation, and $\omega_z$ was median-filtered over a $3\times 3$ stencil to suppress pixel-scale noise. The integral was scaled by the pixel area $\Delta x\,\Delta y$, and $E(t)$ was smoothed with an 11-frame moving average ($\approx 0.01$ s). Doubling or halving the smoothing window did not alter conclusions.  

\subsection{Vortex-core identification and tracking}
The coherent counter-clockwise vortex core was identified using a second-moment method adapted from \citet{steiner2023vortex}.  
A binary mask was defined by thresholding
\begin{equation}
M(x,y) =
  \begin{cases}
    1, & \omega_z(x,y) < \omega_{\text{th}}, \\[6pt]
    0, & \text{otherwise},
  \end{cases}
\end{equation}
with a threshold chosen as $\omega_{\text{th}} = -0.15\,\omega_{z,\max}$ in each frame.  
We verified robustness of this choice by repeating the analysis with threshold coefficients in the range $0.10$–$0.25$.  
Across all cases, the resulting centroid position shifted by less than $0.2\%$ of the cylinder radius, the equivalent radius $r_{\text{eq}}$ changed by less than $1.5\%$, and the circulation $\Gamma$ by less than $2\%$, all of which are well within the PIV uncertainty ($\pm 3\%$).  
The largest connected region was retained, and the vorticity-weighted centroid was computed as
\begin{equation}
x_F = \frac{\sum \tilde{\omega}_z\,x}{\sum \tilde{\omega}_z},
\qquad
y_F = \frac{\sum \tilde{\omega}_z\,y}{\sum \tilde{\omega}_z},
\end{equation}
where $\tilde{\omega}_z$ denotes the thresholded vorticity field.

The second central moments about $(x_F,y_F)$ were evaluated as
\begin{equation}
\mu_{20} = \sum \tilde{\omega}_z (x-x_F)^2, \qquad
\mu_{02} = \sum \tilde{\omega}_z (y-y_F)^2,
\end{equation}
and the equivalent radius defined as
\begin{equation}
r_{\mathrm{eq}} = (\mu_{20}\,\mu_{02})^{1/4}.
\end{equation}
This provides a robust, shape-independent measure of vortex size. The associated equivalent area is $A_r = \pi r_{\mathrm{eq}}^2$.  

Only the counter-clockwise (left-hand) vortex was consistently illuminated, as the laser sheet originated from the left and the cylinder shadowed the opposite side.  

\subsection{Vortex kinematics}
The vortex speed was computed from the centroid trajectory by finite differences,
\begin{equation}
v_{\mathrm{vort}}(t) = \sqrt{\left(\frac{dx_F}{dt}\right)^2 + \left(\frac{dy_F}{dt}\right)^2},
\end{equation}
with smoothing to suppress numerical noise. Both the scalar speed and the components $v_{x,\mathrm{vort}}$ and $v_{y,\mathrm{vort}}$ were analyzed to quantify drift and vertical ascent.  

All histories are reported as functions of the non-dimensional cylinder position $y_{\mathrm{cyl}}/a$, so that comparisons across accelerations are aligned by geometry rather than by time. For clarity, Figure \ref{fig:vortex_speed} additionally employs a semi-logarithmic vertical scale to emphasize the collapse of the normalized vortex-advection speed.

\subsection{Impulse-based scaling}
\label{sec:scaling_theory}
To rationalize the growth in circulation, we employ an impulse balance. The Kelvin impulse $I$ of a compact vortex of core size $\ell = O(a)$ scales as  \[I \sim \rho\,\Gamma\,\ell^2,\] where $\Gamma$ is the vortex circulation and $\rho$ the fluid density  
(see, \citet{Saffman1985}).  

For a cylinder exiting under constant acceleration $\alpha$, the body speed satisfies  \[v_{\mathrm{cyl}}^2 = 2\alpha \,\Delta y_{\mathrm{cyl}},\] with $\Delta y_{\mathrm{cyl}}$ the vertical displacement since release.  The associated timescale is $t \sim \sqrt{\Delta y_{\mathrm{cyl}}/\alpha}$, giving the impulse imparted to the fluid as,  
\begin{equation}
I \sim \rho\,a^2 \sqrt{\alpha\,\Delta y_{\mathrm{cyl}}}.
\label{eq:ImpulseScaling}
\end{equation}

Here, $I$ is the same Kelvin impulse defined above, now estimated from the body motion. Equating this to the vortex impulse $\rho\,\Gamma\,a^2$ yields the scaling law,  
\begin{equation}
\Gamma \sim C_1 \sqrt{\alpha\,\Delta y_{\mathrm{cyl}}^{\,3}} .
\label{eq:GammaScaling}
\end{equation}

As the vortex approaches the free surface, its evolution is modified by interaction with the interface. We represent this phenomenologically with a depth-dependent correction,  
\begin{equation}
\Gamma \sim C_1 \sqrt{\alpha\,\Delta y_{\mathrm{cyl}}^{\,3}}\;
F\!\left(\frac{y_{\mathrm{vort}}}{a}\right),
\qquad
F \to 1 \ \text{for}\ y_{\mathrm{vort}}/a \gg 1,
\label{eq:GammaDepth}
\end{equation}
where $y_{\mathrm{vort}}/a$ is the instantaneous depth of the vortex centroid in units of radius of the cylinder.  This form is consistent with the collapse observed in Figure \ref{fig:circulation}B before free-surface interaction,  and with the systematic departure once $y_{\mathrm{vort}}/a \approx 5$ is reached.

\section{Results and Discussion}
\label{sec:ResultsDiscussion}

In this section, we present the measured evolution of the counter-clockwise starting vortex generated as a circular cylinder leaves the water under constant acceleration. Although a symmetric pair of counter-rotating vortices may form beneath the cylinder, our laser illumination (left-to-right as shown in figure~\ref{PIV_setup}) is blocked by the opaque aluminium body on the right-hand side. Therefore, only the left-hand counterclockwise vortex is consistently illuminated and quantitatively characterized.

\subsection{Vortex size}

\begin{figure}[htbp]
    \centering
    \includegraphics[width=0.75\textwidth]{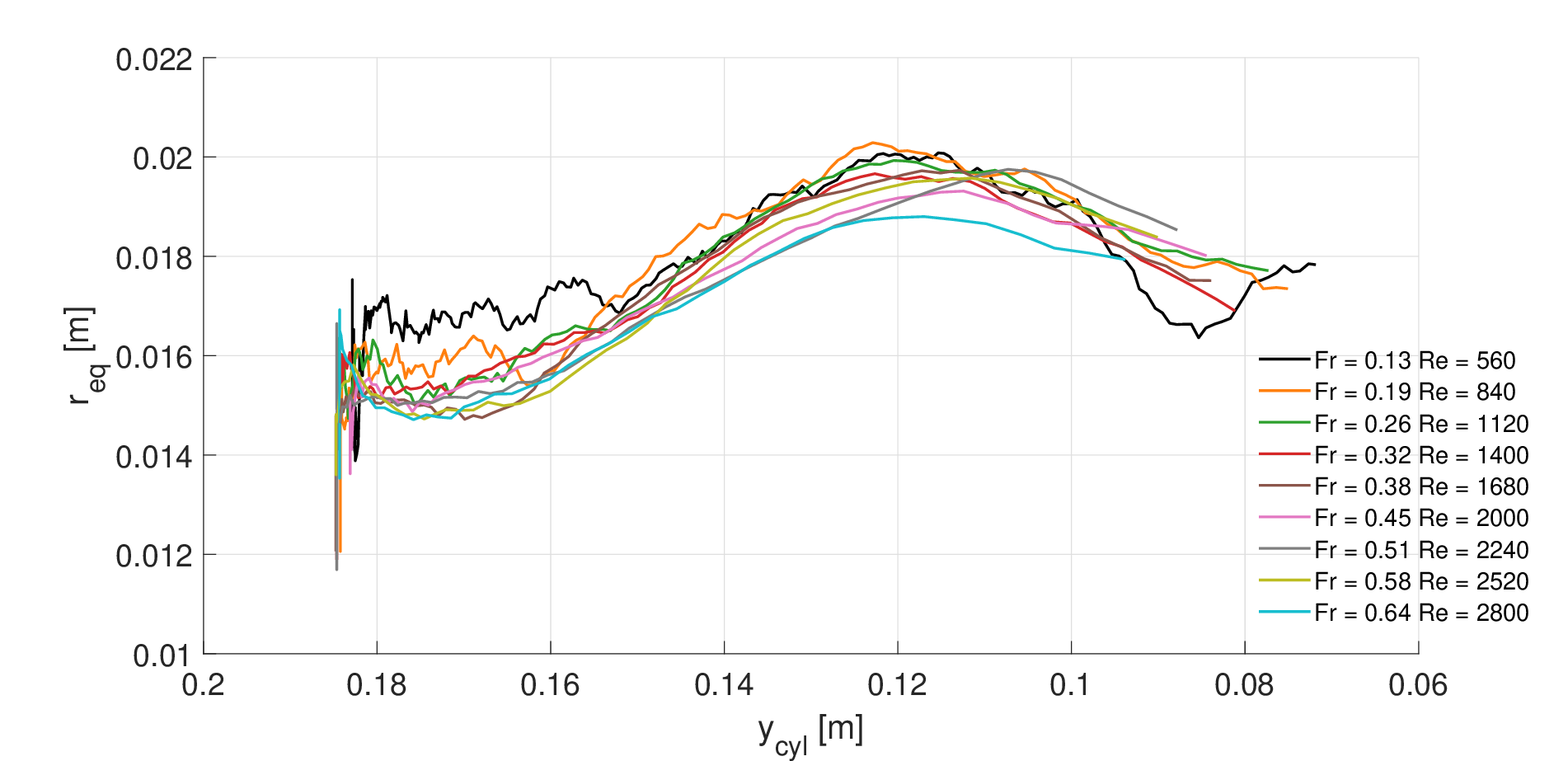}
   \caption{Equivalent vortex radius $r_{\text{eq}} = \sqrt{A/\pi}$ as a function of the 
   cylinder position $y_{cyl}$.}

    \label{fig:req}
\end{figure}

Figure~\ref{fig:req} shows the equivalent radius of the starting vortex, $r_{\mathrm{eq}}=\sqrt{A/\pi}$, obtained from the second spatial moments of the spanwise vorticity field. 

Right after the core first appears ($y_{\text{cyl}}^{\ast}\!\approx\!10$), $r_{\mathrm{eq}}$ drops to a shallow minimum—an effect attributed to the early re-organisation of the newly separated shear layer—before increasing steadily for the remainder of the record. Between $y_{\text{cyl}}^{\ast}\!\approx\!9$ and $y_{\text{cyl}}^{\ast}\!\approx\!2$, $r_{\mathrm{eq}}$ grows from roughly $1.0\times10^{-2}$ to $1.35\times10^{-2}\,\mathrm{m}$, an increase of about $30\%$. Acceleration shifts the absolute values only modestly: the highest $Fr$ case finishes no more than 15–20\% above the lowest, and all curves remain clustered within that band over the full ascent.

A reasonable working hypothesis is that the continued entrainment of the attached shear layer increases the footprint of the vortex throughout the observed window; no post-peak contraction is evident, so any later balance between axial stretching, viscous diffusion, and residual entrainment must occur below $y_{\text{cyl}}^{\ast}\!\approx\!2$, outside the present field of view. 

Because the out-of-plane strain $\partial w / \partial z$ (where $w$ is the velocity component along the $z$-axis) is not measured, this interpretation remains tentative. It could be tested by volumetric PIV or by a validated three-dimensional simulation.

For consistency and to avoid redundancy, the equivalent radius $r_{\mathrm{eq}}$ is retained as the primary size metric in the remainder of this paper.

\subsection{Internal vortex dynamics}

\begin{figure}[H]
    \centering
    \includegraphics[width=0.75\textwidth]{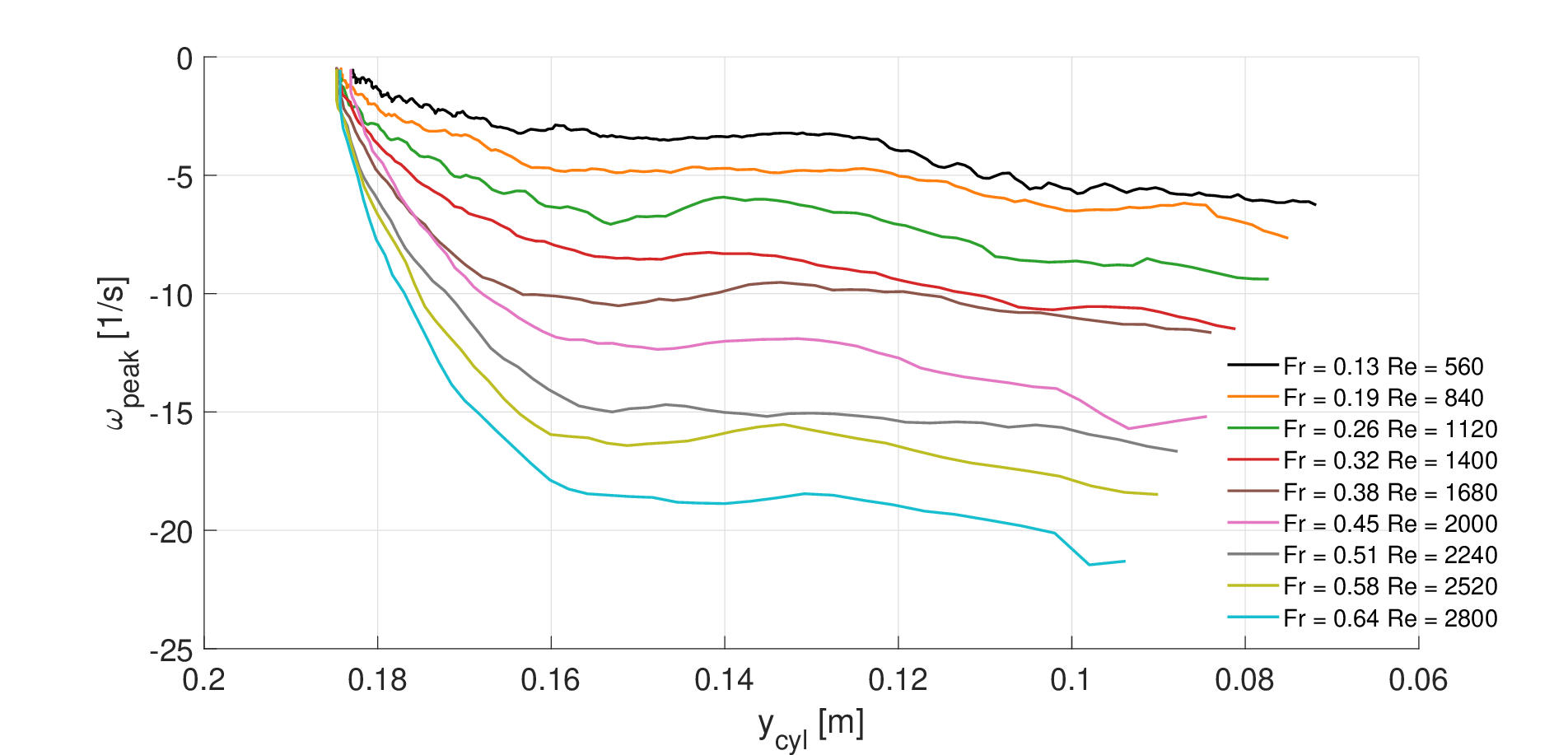}
    \caption{Peak vorticity $\omega_{\text{peak}}$ inside the vortex core as a function of the cylinder position $y_{cyl}$. A sharp rise occurs during the early roll-up ($y_{cyl}^*\!\approx\!10$), followed by gradual decay.}
    \label{fig:omega}
\end{figure}

\begin{figure}[H]
    \centering
    \includegraphics[width=0.75\textwidth]{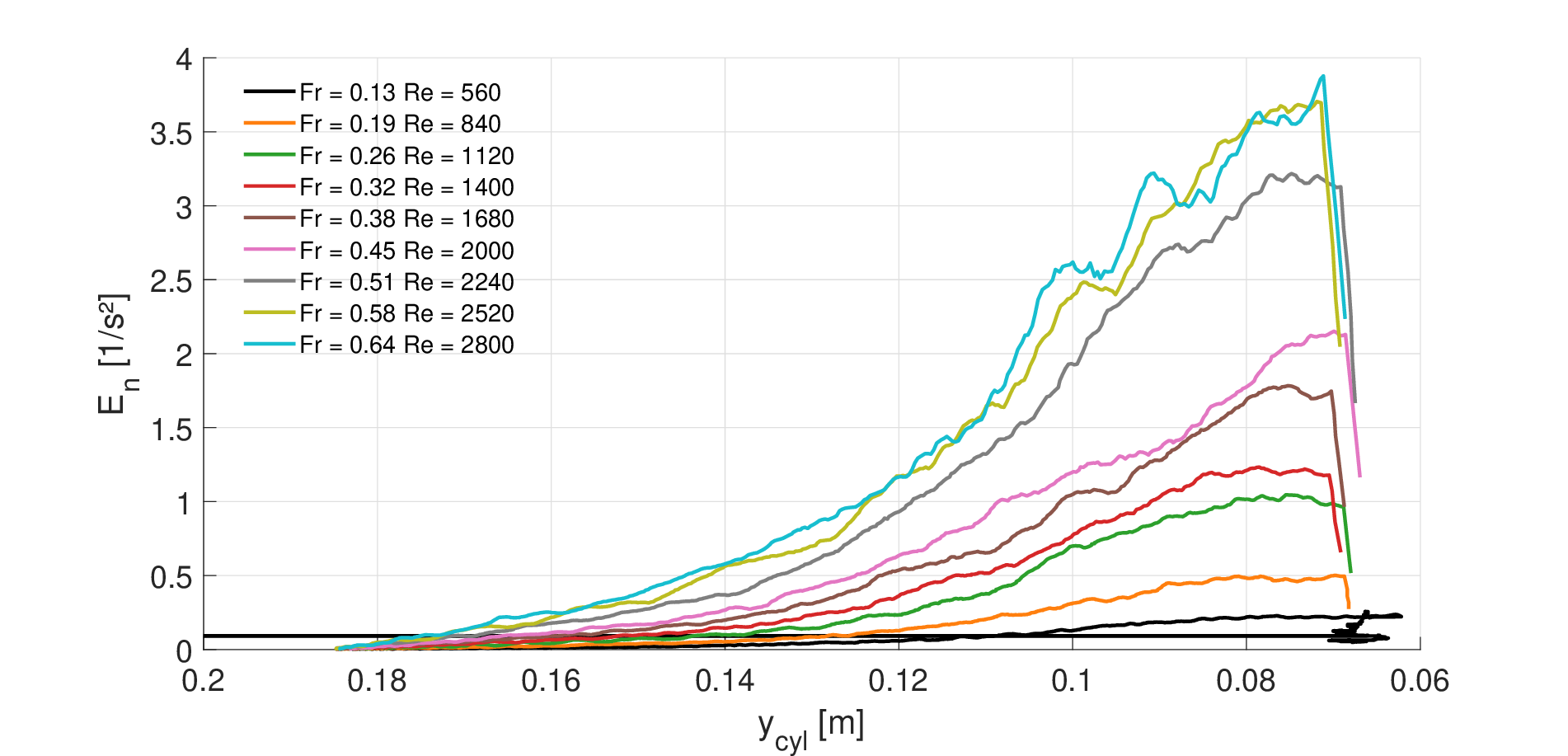}
    \caption{Enstrophy $E_n = \int_\Omega \omega_z^2 \, dA$ vs.\ cylinder position $y_{cyl}$. This measure integrates the squared vorticity field, reflecting vortex intensity and compactness.}
    \label{fig:enstrophy}
\end{figure}

Figure~\ref{fig:omega} shows the evolution of the most negative spanwise vorticity inside the counter-clockwise core, while Figure~\ref{fig:enstrophy} presents the corresponding domain-integrated enstrophy \(E_n=\iint_{\Omega}\omega_z^2\,dA\).

Across all accelerations, three repeatable stages are observed. (1) At larger cylinder heights \(y_{\mathrm{cyl}}\) (early times), the magnitude of \(\omega_{\text{pk}}\) increases rapidly, marking the roll-up phase in which the separated shear layer feeds like-signed vorticity into a compact core. (2) At intermediate \(y_{\mathrm{cyl}}\), this growth halts or slightly reverses, consistent with a lengthening feeding layer and the onset of diffusion together with incipient axial stretching, which begin to offset further intensification. (3) At smaller \(y_{\mathrm{cyl}}\) (closer to the free surface), \(|\omega_{\text{pk}}|\) strengthens again, most clearly at higher \(Fr\); in this regime the shear layer has detached and axial stretching of the isolated vortex tube enhances the local vorticity. The peak values increase systematically with acceleration (e.g. \(\omega_{\text{pk}}\approx -27~\mathrm{s}^{-1}\) at \(Fr_\alpha=0.64\) versus \(\approx -7~\mathrm{s}^{-1}\) at \(Fr_\alpha=0.13\)).

In contrast, \(E_n\) grows monotonically throughout the record and increases as \(y_{\mathrm{cyl}}\) decreases. A simple scaling can rationalize this behavior. If the core contains most of the like-signed vorticity, with an effective area \(A_{\rm eff}\sim \pi r_{\mathrm{eq}}^{2}\) and a characteristic level \(\overline{\omega}\), then
\[\Gamma \;\sim\; \overline{\omega}\,A_{\rm eff}, \qquad E_n \;\sim\; \overline{\omega}^{\,2}\,A_{\rm eff}.\] Eliminating \(\overline{\omega}\) gives
\begin{equation}
E_n \;\sim\; \frac{\Gamma^{2}}{A_{\rm eff}}
\;\sim\; \frac{\Gamma^{2}}{\pi r_{\mathrm{eq}}^{2}}.
\label{eq:EnScaling}
\end{equation}
Since \(\Gamma\) follows the impulse-based growth \(\Gamma \sim \sqrt{\alpha\,\Delta y_{\mathrm{cyl}}^{\,3}}\) (Section~\ref{sec:global_circ}) while \(r_{\mathrm{eq}}\) increases only moderately, Eq.~\eqref{eq:EnScaling} predicts a monotonic, and increasingly steep, increase of \(E_n\) with ascent, in agreement with Fig.~\ref{fig:enstrophy}. The stronger acceleration dependence at higher \(Fr\) then follows directly from the \(\Gamma^{2}\) factor.

Together, these trends show that acceleration chiefly amplifies the intensity of vortex-core dynamics. Peak vorticity responds in a non-monotonic sequence: roll-up, plateau, then renewed growth, whereas enstrophy accumulates steadily, underscoring the persistent role of axial stretching and continued vorticity influx in energizing the starting vortex during water exit.

\subsection{Global circulation}
\label{sec:global_circ}

The internal measures mentioned above revealed an explicit dependence on acceleration. We now turn to the global circulation, which provides an integrated view of the vortex dynamics and allows direct comparison with the impulse-based scaling theory.

The circulation is obtained from the spanwise vorticity field as 
\(\Gamma=\iint_{\Omega}\omega_z\,dA\) (Appendix~B).  The scaling arguments of Section~\ref{sec:scaling_theory} predict that, before surface interaction, circulation growth should follow the impulse-based form in Eq.~\eqref{eq:GammaScaling}, with deviations described by Eq.~\eqref{eq:GammaDepth} once the vortex approaches the free surface.

\begin{figure}[H]
    \centering
    \includegraphics[width=1\textwidth]{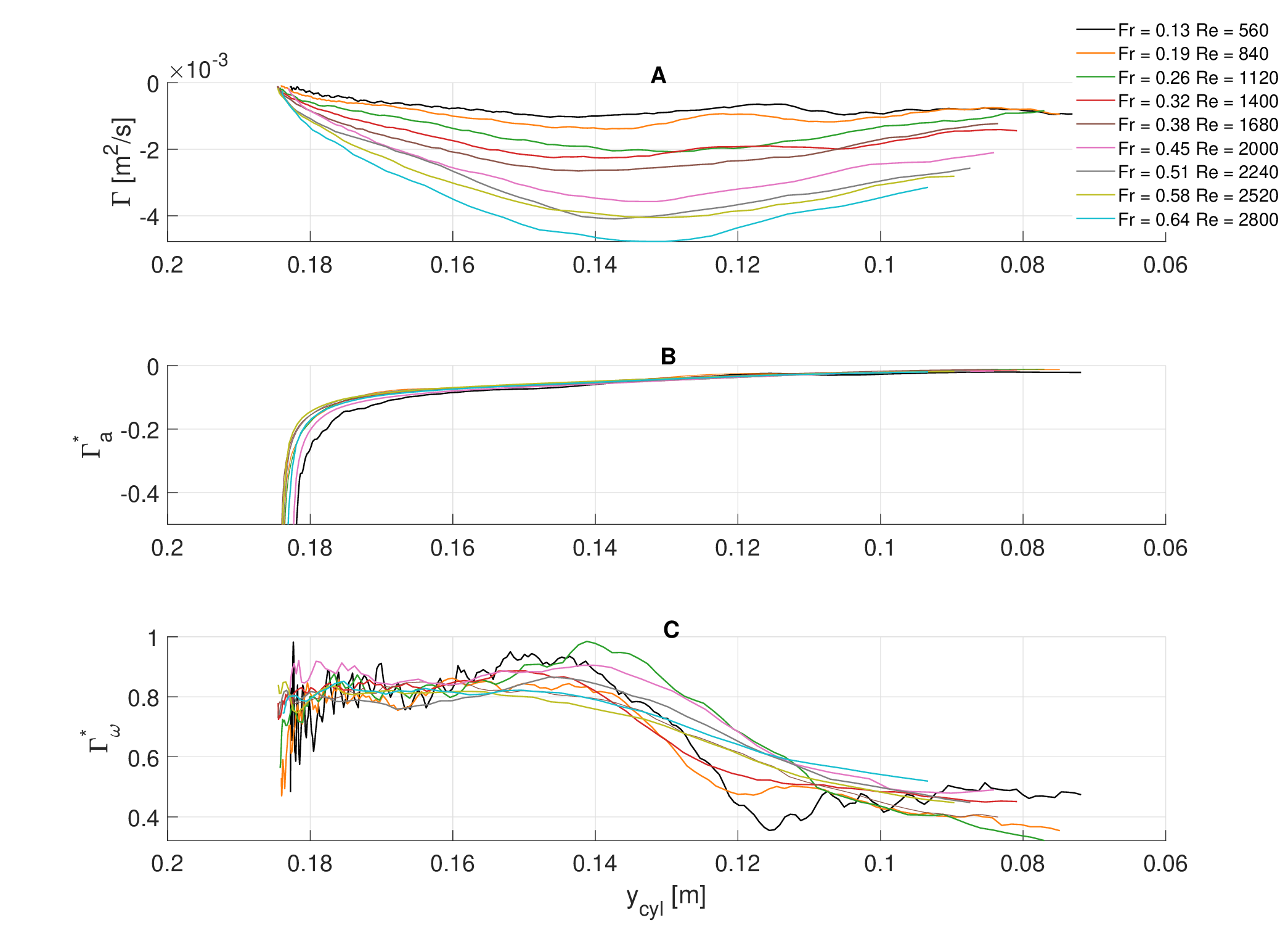}
     \caption{Circulation scaling with cylinder position $y_{cyl}$. (A) Raw circulation $\Gamma$; ((B) Acceleration-scaled \( \Gamma_a^{\ast} = \Gamma / \sqrt{\alpha (y_{\mathrm{cyl}}-y_{cyl(0)})^3} \),  where \(y_{cyl(0)}\) is the initial immersion depth ; (C) vorticity-scaled $\Gamma_\omega^* = \Gamma / (\omega_{\text{peak}} \pi r_{\text{eq}}^2)$. Both scalings collapse data across accelerations, supporting their robustness.}
    \label{fig:circulation}
\end{figure}

Figure~\ref{fig:circulation}(a) shows the dimensional circulation \(\Gamma(t)\).Across all accelerations, \(\Gamma\) increases nearly linearly until the cylinder reaches \(y_{\mathrm{cyl}}\approx 0.04\;\mathrm{m}\), after which it levels off. In the highest-\(Fr_\alpha\) case, a slight decrease (\(\sim 5\%\)) appears after the peak, which may reflect early viscous diffusion or partial cancellation by opposite-signed vorticity.

Figure~\ref{fig:circulation}(b) demonstrates that normalizing circulation with the impulse form,
\[\Gamma_{a}^{\ast}(t) \;=\; \frac{\Gamma(t)}{\sqrt{\alpha\,\Delta y_{\text{cyl}}^{\,3}}},\qquad 
\Delta y_{\text{cyl}} = y_{\text{cyl}}(t)-y_{\text{cyl}}(0),\]
reduces all nine acceleration cases to a single curve. The residual scatter remains within the measurement uncertainty (\(\pm 5\%\)), confirming that circulation growth is governed predominantly by the acceleration impulse, as anticipated by the model.

For comparison, Fig.~\ref{fig:circulation}(c) applies a core-based normalization,\[\Gamma_{\omega}^{\ast}(t) \;=\; \frac{\Gamma(t)}{\omega_{\text{pk}}(t)\,\pi r_{\text{eq}}^{2}(t)}.\] This scaling collapses the early growth, reflecting the proportionality between 
\(\Gamma\) and \(\omega_{\text{pk}} r_{\text{eq}}^{2}\) while entrainment remains modest. However, as the core expands and redistributes vorticity,
\(\omega_{\text{pk}}\) decreases while \(r_{\text{eq}}\) continues to grow, and the curves diverge 20–25\%.  Thus, the core scaling captures only the initial phase, whereas the impulse scaling provides similarity across the full ascent, consistent with the minimal model.

The success of impulse-based scaling arises because circulation is fundamentally governed by the hydrodynamic impulse imparted to the fluid, which depends only on the acceleration history of the body. In this formulation, vortex circulation represents the most direct transfer of momentum from the accelerating cylinder to the surrounding fluid, and therefore collapses naturally when normalized by the impulse budget.

In contrast, core-based normalization presumes that the vortex remains compact and that essentially all circulation is confined within the region characterized by \(\omega_{\text{pk}}\) and \(r_{\text{eq}}\). This assumption holds only in the earliest stages of roll-up, when the vorticity distribution is still localized. However, as the ascent progresses, vorticity is progressively shed into surrounding shear layers, entrained into secondary motions, and redistributed by stretching and diffusion. Once this redistribution occurs, the core measures no longer represent the entire impulse content, and the collapse breaks down across accelerations.

The comparison thus highlights a fundamental distinction: impulse scaling is tied to the \emph{forcing} imposed by the body motion, whereas core-based scaling reflects the \emph{internal response} of the evolving vortex. Only the former provides a universal description because it remains insensitive to the subsequent redistribution of vorticity. In this sense, the impulse form captures the true forcing–response link in starting-vortex dynamics, while core-based scalings are inherently restricted to the very early, compact-core regime.

Having established how circulation grows under acceleration and why impulse scaling provides the correct collapse, we now examine how the vortex itself is transported. The circulation budget determines the strength of the structure, but its ability to interact with the free surface also depends on the trajectory and convection speed of the core relative to the cylinder. Accordingly, we next characterize the translation of the vortex.

\subsection{Vortex translation}

\begin{figure}[htbp]
    \centering
    \begin{subfigure}[b]{0.75\textwidth}
        \centering
        \includegraphics[width=\textwidth]{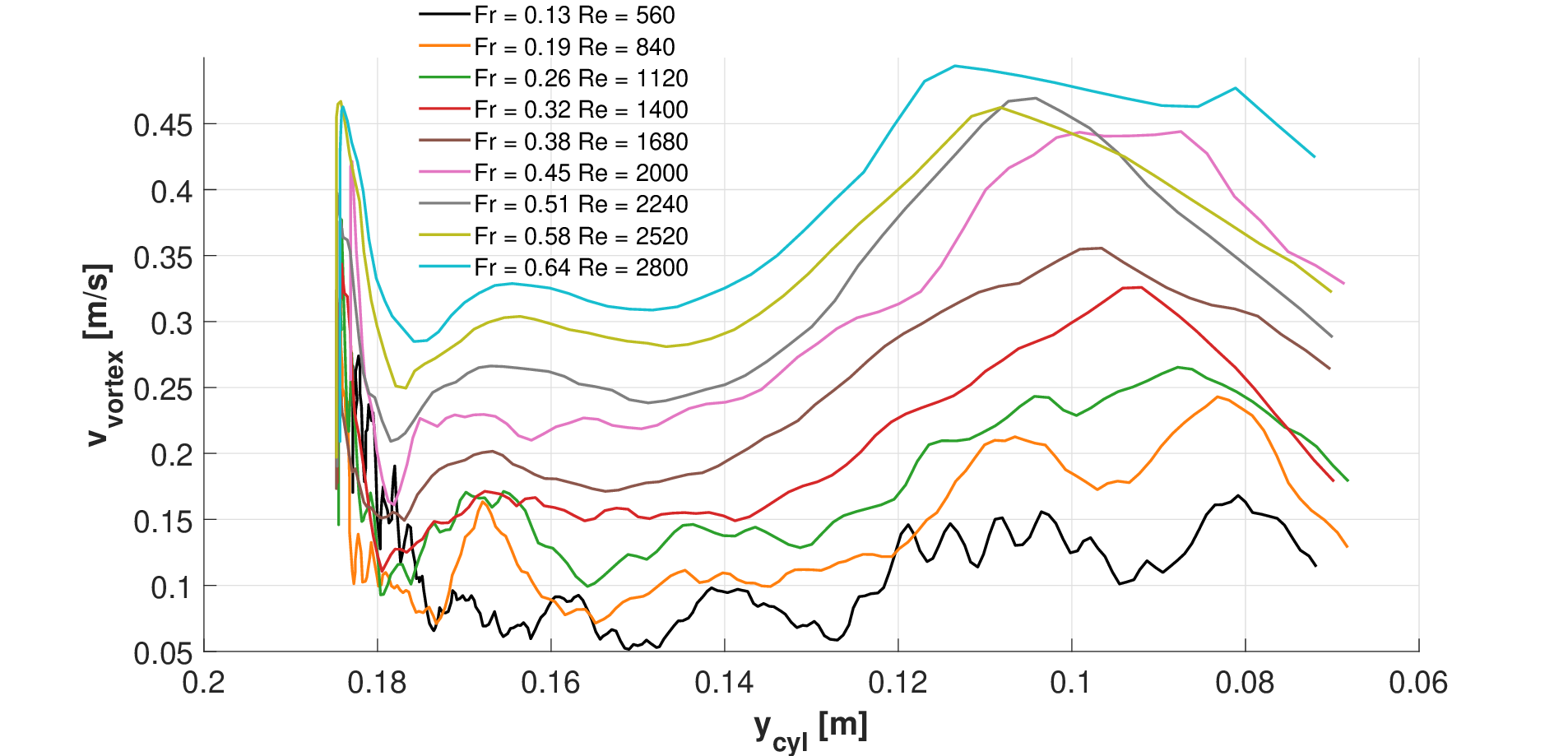}
        \caption{Absolute vortex advection speed as a function of the
       actual  cylinder center position $y_{cyl}$.}
        \label{fig:vortex_speed_a}
    \end{subfigure}

    \vspace{0.5em} % spacing between panels

    \begin{subfigure}[b]{0.75\textwidth}
        \centering
        \includegraphics[width=\textwidth]{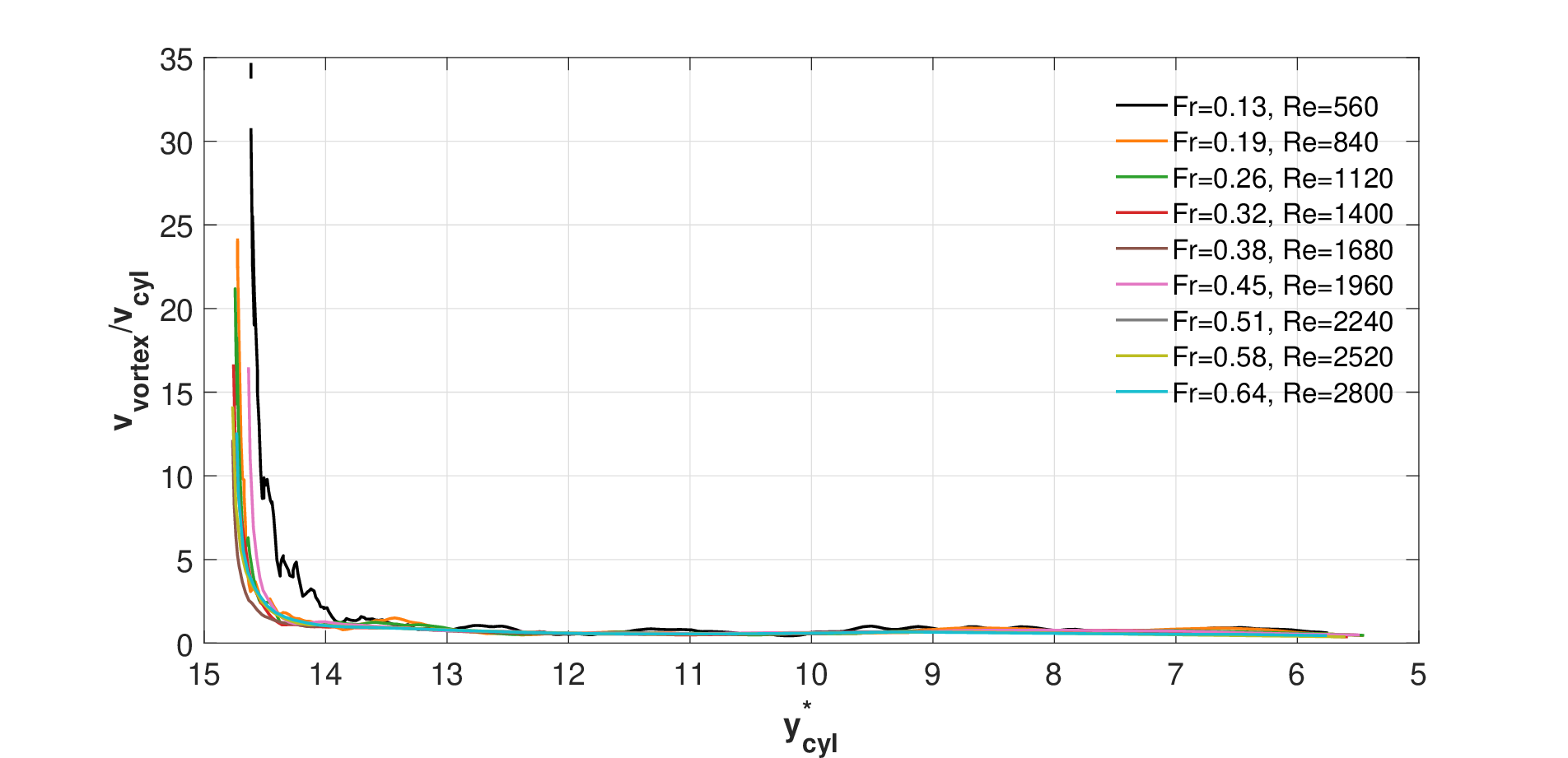}
        \caption{Vortex advection speed normalized by the instantaneous cylinder
        velocity $v_{\mathrm{cyl}}=\sqrt{2\alpha \,\Delta y_{\mathrm{cyl}}}$ as a function of the
       normalized  cylinder center position $y_{cyl}^*$.}
        \label{fig:vortex_speed_b}
    \end{subfigure}
   \caption{Vertical advection speed of the vortex core. (A) Absolute vortex advection speed vs.\ cylinder position $y_{cyl}$; (B) normalized vortex advection speed, scaled by the instantaneous body velocity $v_{\mathrm{cyl}}=\sqrt{2\alpha \,\Delta y_{\mathrm{cyl}}}$, vs.\ normalized position $y_{cyl}^*$.}
    \label{fig:vortex_speed}
\end{figure}

Figure~\ref{fig:vortex_speed}a shows how the circulation-carrying structure is 
advected upward: the absolute core speed is plotted versus $y_{\mathrm{cyl}}$. 
When normalized by the instantaneous body speed $v_{\mathrm{cyl}}=\sqrt{2\alpha\,\Delta y_{\mathrm{cyl}}}$, the curves collapse (Fig.~\ref{fig:vortex_speed}b), indicating that the vortex is transported at a nearly constant fraction of the body speed across all accelerations.

\begin{figure}[htbp]
    \centering
    \includegraphics[width=0.5\textwidth]{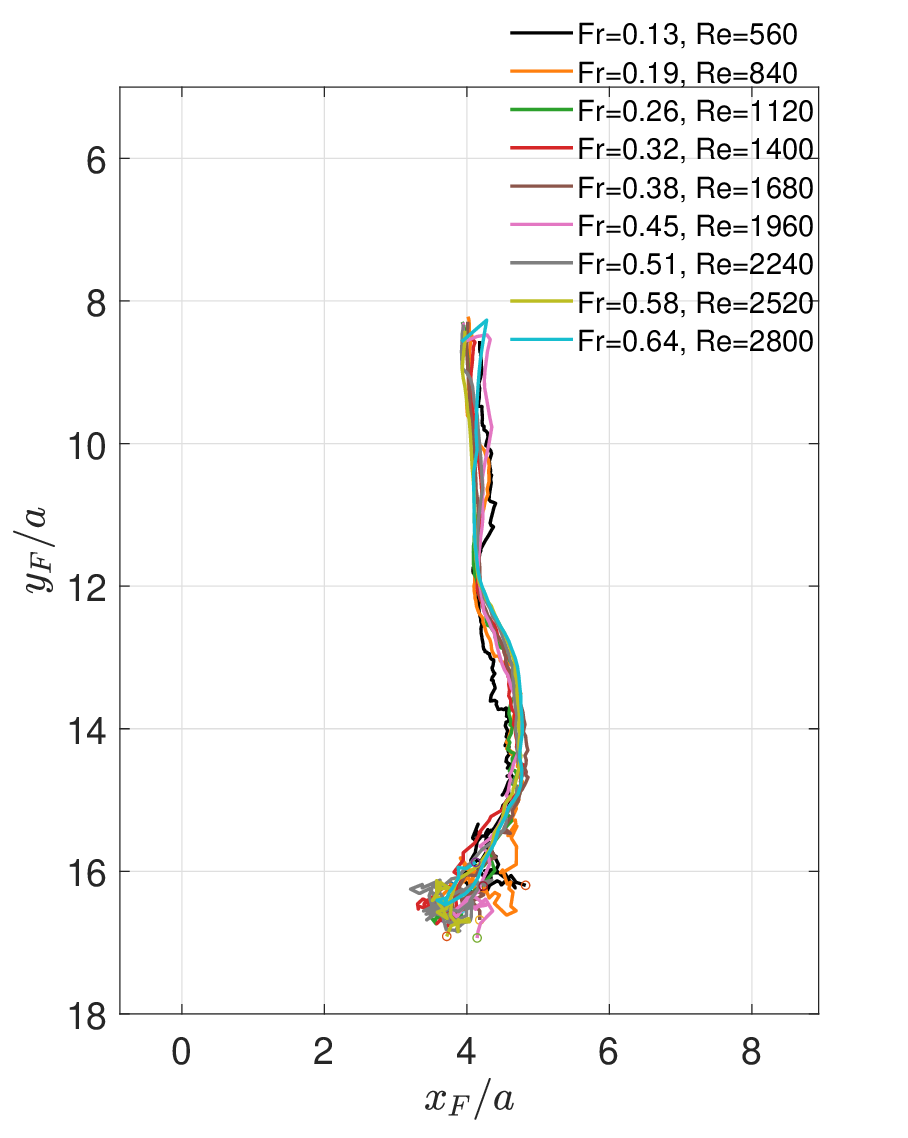}
    \caption{Non-dimensional vortex-centroid trajectory $(x_F/a,\,y_F/a)$ during the exit event. Here, $y_{c,\mathrm{vort}}^* \equiv y_F/a$ and $x_F/a$ denote the vertical and horizontal positions of the vortex centroid, respectively. Lateral offsets remain below $0.2a$.}
    \label{fig:trajectory}
\end{figure}

Non-dimensional trajectories $(x_F/a,\,y_F/a)$ in Fig.~\ref{fig:trajectory} are nearly vertical, with lateral offsets $<0.2a$ for all cases. Scaling by $a$ also reveals a consistent onset of free-surface interaction near $y_{c,\mathrm{vort}}^*\approx 5$, independent of acceleration.

%\begin{figure}[H]
%\    \centering
 %\   \includegraphics[width=1\textwidth]{Areaa.eps}
 %\   \caption{Vortex area \( A_r \) as a function of vortex centroid position \( y_{cyl} \) for various \( Fr \) and \( Re \).}
%\    \label{fig:area}
%\\end{figure}

\section{Conclusions}
\label{sec:conclusions}

We have investigated the starting vortex generated as a circular cylinder exits quiescent water under constant acceleration. Time-resolved PIV measurements revealed that circulation grows rapidly during shear-layer attachment and then levels off once detachment occurs, whereas enstrophy accumulates monotonically throughout the ascent. Vortex-core diagnostics such as peak vorticity and equivalent radius reflect the redistribution of vorticity within the core, but do not yield a consistent collapse across accelerations.

By contrast, circulation normalized with the impulse-based form,
\[\Gamma_a^{\ast} = \frac{\Gamma}{\sqrt{\alpha\,\Delta y_{\mathrm{cyl}}^{3}}},\] collapsed all accelerations within experimental uncertainty. This demonstrates that vortex strength is controlled by the hydrodynamic impulse imparted by the accelerating body. The result underscores a fundamental distinction: forcing-based scalings remain universal once vorticity has been shed into the wake, whereas core-based normalisations succeed only in the earliest compact-core regime. The vortex trajectory further revealed predominantly vertical advection at a fixed fraction of the body speed, with lateral offsets below $0.2a$.

Taken together, these findings establish impulse-based scaling as a robust framework for predicting vortex formation in water-exit flows and provide quantitative benchmarks for validating free-surface CFD. Beyond benchmarking, the results suggest design principles for unsteady propulsion and recovery manoeuvres where momentum transfer is dominated by hydrodynamic impulse. Future work should extend these measurements to three-dimensional geometries, variable accelerations, and wave-modulated exits to test the generality of the impulse-scaling framework.

While the present study provides a quantitative framework for vortex dynamics during the water exit of a circular cylinder under constant acceleration, several important limitations remain. The experiments are restricted to planar measurements and to a single, rigid geometry in quiescent fluid, which may not capture the full complexity of real-world configurations. Future research should explore the effects of three-dimensional flow structures, varied body geometries (such as spheres, plates, or irregular shapes), orientation angles, and material flexibility. Additionally, introducing environmental disturbances, such as background turbulence or surface waves, would provide a more comprehensive understanding of vortex evolution and impulse transfer. Extending these measurements and scaling analyses to new geometries and flow conditions is essential not only to assess the universality of the impulse-based framework but also to refine predictive models for engineering applications involving unsteady water-exit phenomena.

\backmatter

\section*{Acknowledgements}

The financial support of the Belgian Fund for Scientific Research under research project WOLFLOW (F.R.S.-FNRS, PDR T.0021.18) is gratefully acknowledged. Part of the experimental setup was financed by {\it Fonds Sp\'eciaux} from ULi\`ege. SD is F.R.S--FNRS Senior Research Associate

\section*{Authors Contribution}

IA: Designed and conducted the experiments, processed the PIV data, developed the vortex identification algorithms, and drafted the original manuscript and revised the manuscript. 

SD: Supervised the overall research project, provided critical feedback on experimental design and analysis, contributed to the discussion and conclusion sections, and revised the manuscript. 
\begin{appendices}
\appendix
\section{Cylinder Kinematics and Validation of Constant Acceleration}
\label{Cylinder kinematics}
The prescribed vertical motion of the cylinder is
\begin{equation}
  y_{\mathrm{cyl}}(t)
  = y_0 + v_0\,t + \tfrac12\,\alpha_{\mathrm{cmd}}\,t^2,
  \label{eq:cylMotion}
\end{equation}
with $y_0$, $v_0$ the initial position and velocity, and $a_{\mathrm{cmd}}$ the commanded constant acceleration.  Differentiation gives
\begin{equation}
  v_{\mathrm{cyl}}(t)
  = \dot y_{\mathrm{cyl}}(t)
  = v_0 + \alpha_{\mathrm{cmd}}\,t,
  \label{eq:cylVelocity}
\end{equation}
\begin{equation}
  a_{\mathrm{cyl}}(t)
  = \ddot y_{\mathrm{cyl}}(t)
  = \alpha_{\mathrm{cmd}}.
  \label{eq:cylAccel}
\end{equation}

To verify constant acceleration experimentally, we proceed as follows:

\begin{enumerate}
  \item Extract the position time‐series \( \{t_i,\,y_{\rm cyl}(t_i)\}\).
  \item Compute the instantaneous acceleration via finite differences and smoothing:
    \begin{equation}
      \alpha_{\rm est}(t_i)
      = \frac{y_{\rm cyl}(t_{i+1}) - 2\,y_{\rm cyl}(t_i) + y_{\rm cyl}(t_{i-1})}{\Delta t^2},
      \label{eq:accFiniteDiff}
    \end{equation}
    then apply a moving‐average filter to reduce noise.
  \item Identify the contiguous interval where 
    \(\bigl|\alpha_{\rm est}(t)-\alpha_{\rm cmd}\bigr|\le5\%\), ensuring we analyze only the truly constant‐acceleration portion.
  \item Perform a robust quadratic fit
    \begin{equation}
      \hat y(t)=\hat y_0 + \hat v_0\,t + \tfrac12\,\hat \alpha\,t^2
      \label{eq:quadFitSimple}
    \end{equation}
    over that interval, extracting the fitted acceleration \(\hat a\).
  \item Evaluate the fit’s \(R^2\) and the residual standard deviation of the acceleration, and plot \(a_{\rm est}(t)\) alongside the commanded \(a_{\rm cmd}\) and the constant fit \(\hat a\).
\end{enumerate}

\begin{figure}[ht]
  \centering
  \includegraphics[width=0.8\textwidth]{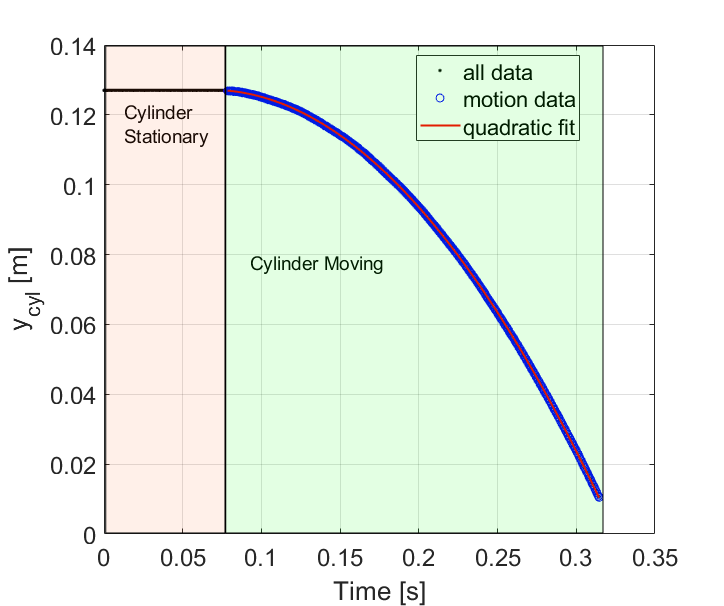}
  \caption{Validation of constant‐acceleration cylinder motion. The orange region indicates when the cylinder is stationary, and the green region indicates when the cylinder is moving with constant acceleration of $4$ $m/s^2$.  Instantaneous acceleration \(\alpha_{\rm est}(t)\) (black dots), commanded
    value \(\alpha_{\rm cmd}\) (blue dashed line), and robust quadratic fit
    \(\widehat\alpha\) (red line).  The black points remain within \(\pm5\%\) of
    \(\alpha_{\rm cmd}\), and the fitted \(\widehat\alpha\) agrees to within a few percent,
    confirming constant‐acceleration kinematics.}
  \label{fig:accel_validation}
\end{figure}

Figure~\ref{fig:accel_validation} shows the vertical position of the cylinder, \(y_{\mathrm{cyl}}(t)\), during the upward motion, along with a quadratic fit (red curve) of the form \(y(t) = y_0 + v_0\,t + \tfrac{1}{2} \hat a\, t^2\). This fit is restricted to the portion of the trajectory corresponding to constant acceleration, prior to the transition to constant velocity and eventual deceleration. The fitted acceleration \(\hat a\) closely matches the commanded value \(a_{\mathrm{cmd}}\), with residuals typically below 0.1~mm and coefficient of determination \(R^2 \approx 1\). This confirms that the cylinder underwent the intended constant-acceleration motion during the analyzed phase.

%-----------------------------------------------------------------
%  Velocity reconstruction, vortex detection, and second–moment quantification
%-----------------------------------------------------------------
\section{Velocity reconstruction, vortex detection, and second–moment quantification}
\renewcommand{\theequation}{B\arabic{equation}}
\setcounter{equation}{0}
\label{sec:vortex_processing_chain}

The dominant counter-clockwise starting vortex in every PIV frame is
identified with the second–moment method of \citet{steiner2023vortex},
which requires no a priori shape assumption and is largely insensitive
to threshold choice.

%---------------------------
\subsection*{Pre-processing}
A Gaussian filter (\(\sigma=2\)~pixels) is applied to each instantaneous
velocity field \(U(x,y),V(x,y)\) to suppress camera noise.
Span-wise vorticity is then obtained by centred differences,
\begin{equation}
\omega_z(x,y)=\frac{\partial V}{\partial x}-\frac{\partial U}{\partial y}.
\end{equation}

%---------------------------
\subsection*{Thresholding and region selection}
Pixels with strong negative vorticity are retained via
\begin{equation}
\omega_{\text{th}}=-0.15\,\omega_{z,\max}(t),\qquad
M(x,y)=
\begin{cases}
1,&\omega_z(x,y)<\omega_{\text{th}},\\
0,&\text{otherwise}.
\end{cases}
\end{equation}

To check the influence of the threshold coefficient, we repeated the entire processing chain with \(\alpha=0.10,\;0.12,\;0.18,\;0.20,\;0.25\). Across 300 frames, the vortex centroid shifted by \(<0.15\,\text{px}\) ( \(<0.2\%\) of \(R\) ),
\(r_{\text{eq}}\) changed by \(<1.5\%\), and \(\Gamma\) by \(<2.1\%\).
Because these variations lie well inside the PIV uncertainty (\(\pm 3\%\). we retain the mid-range value \(\alpha=0.15\) for all results presented in Section~\ref{sec:ResultsDiscussion}.

The largest four-connected component of the binary mask \(M\) defines
the vortex region \(\mathcal{R}\) containing \(N\) pixels
\((x_i,y_i)\) with values \(\omega_i=\omega_z(x_i,y_i)\).

%---------------------------
\subsection*{Centroid and size from second moments}
The vorticity-weighted centroid is
\begin{equation}
x_F=\frac{\sum_{i=1}^N\omega_i x_i}{\sum_{i=1}^N\omega_i}, \qquad
y_F=\frac{\sum_{i=1}^N\omega_i y_i}{\sum_{i=1}^N\omega_i}.
\end{equation}
Central second moments about \((x_F,y_F)\) are
\begin{equation}
\mu_{20}=\sum_{i=1}^N\omega_i (x_i-x_F)^2, \qquad
\mu_{02}=\sum_{i=1}^N\omega_i (y_i-y_F)^2 .
\end{equation}
An equivalent radius that preserves these moments is
\begin{equation}
r_{\text{eq}}=(\mu_{20}\mu_{02})^{1/4},
\end{equation}
and the corresponding area is \(A_r=\pi r_{\text{eq}}^{2}\).

%---------------------------
\subsection*{Circulation and normalisation}
Circulation over \(\mathcal{R}\) is
\begin{equation}
\Gamma=\iint_{\mathcal{R}}\omega_z\,dA .
\end{equation}
Two non-dimensional forms are used in the results:
\begin{align}
\Gamma_a^{\ast}&=\frac{\Gamma}{\sqrt{\alpha\,\Delta y_{cyl}^{\,3}}}, \\
\Gamma_\omega^{\ast}&=\frac{\Gamma}{\omega_{\text{pk}}\pi r_{\text{eq}}^{2}},
\end{align}
where \(\alpha\) is the imposed acceleration, \(\Delta y_{cyl}\) the
centre-of-mass rise since \(t=0\), and
\(\omega_{\text{pk}}=\min_{\mathcal{R}}\omega_z\).

%---------------------------
\subsection*{Post-processing}
The centroid trace \((x_F(t),y_F(t))\) is smoothed with an 11-frame moving average to suppress sub-pixel jitter. Time series of \(\Gamma(t)\), \(r_{\text{eq}}(t)\), \(\omega_{\text{pk}}(t)\) and enstrophy \(E(t)\) retain full temporal resolution and form the data set analysed in

\begin{figure}
    \centering
    \includegraphics[width=1\textwidth]{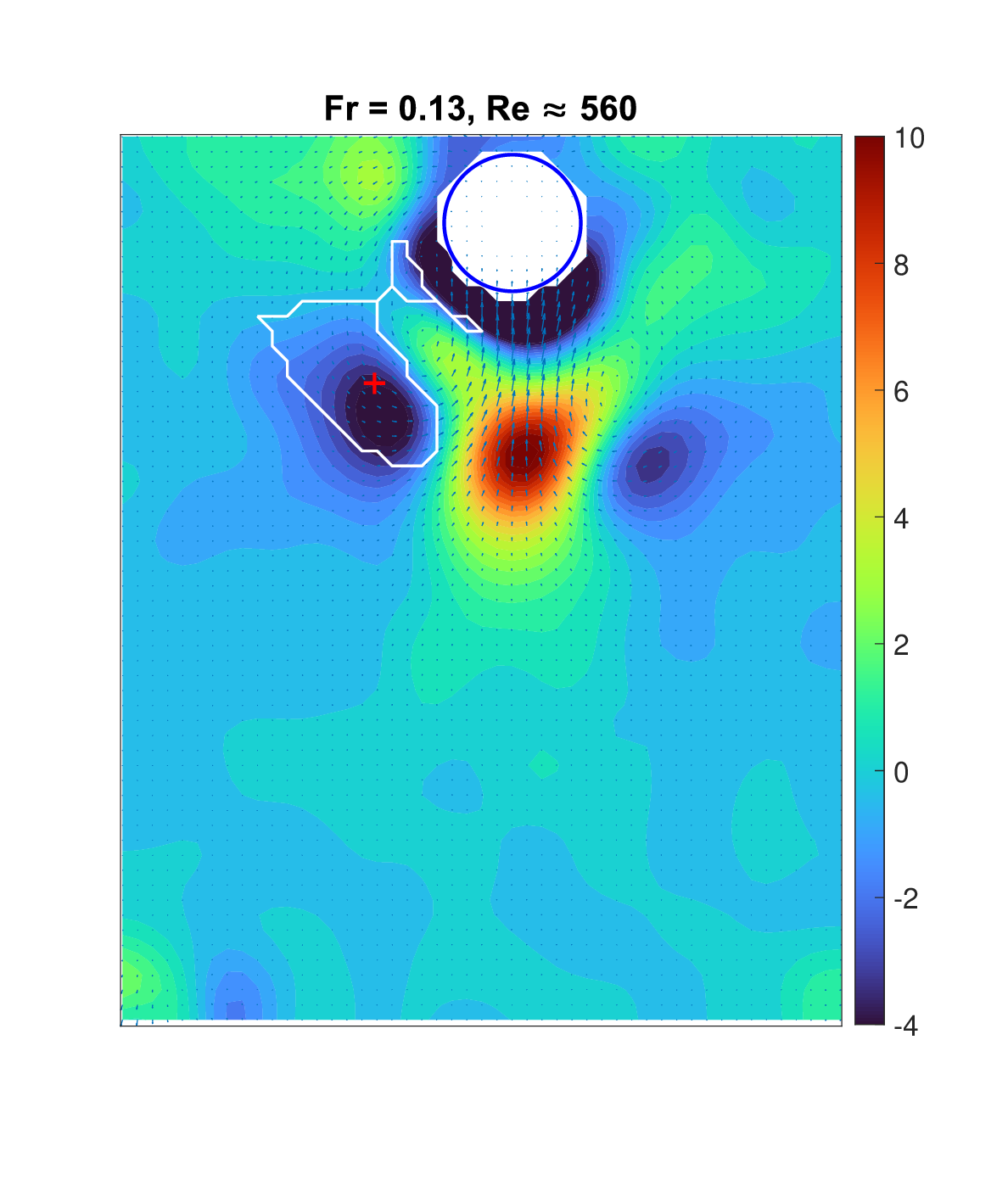}
    \caption{Identification of vortex at $y_{cyl}^* = 1$ for Fr =0.13, Re $\equiv$ $560$. The white boundary shows the vortex-identified region. Vorticity field example. Darker shading corresponds to stronger vorticity magnitudes. The black spot beneath the cylinder is an imaging shadow and not a physical vortex.}
    \label{fig:vortex1}
\end{figure}

\begin{figure}
    \centering
    \includegraphics[width=1\textwidth]{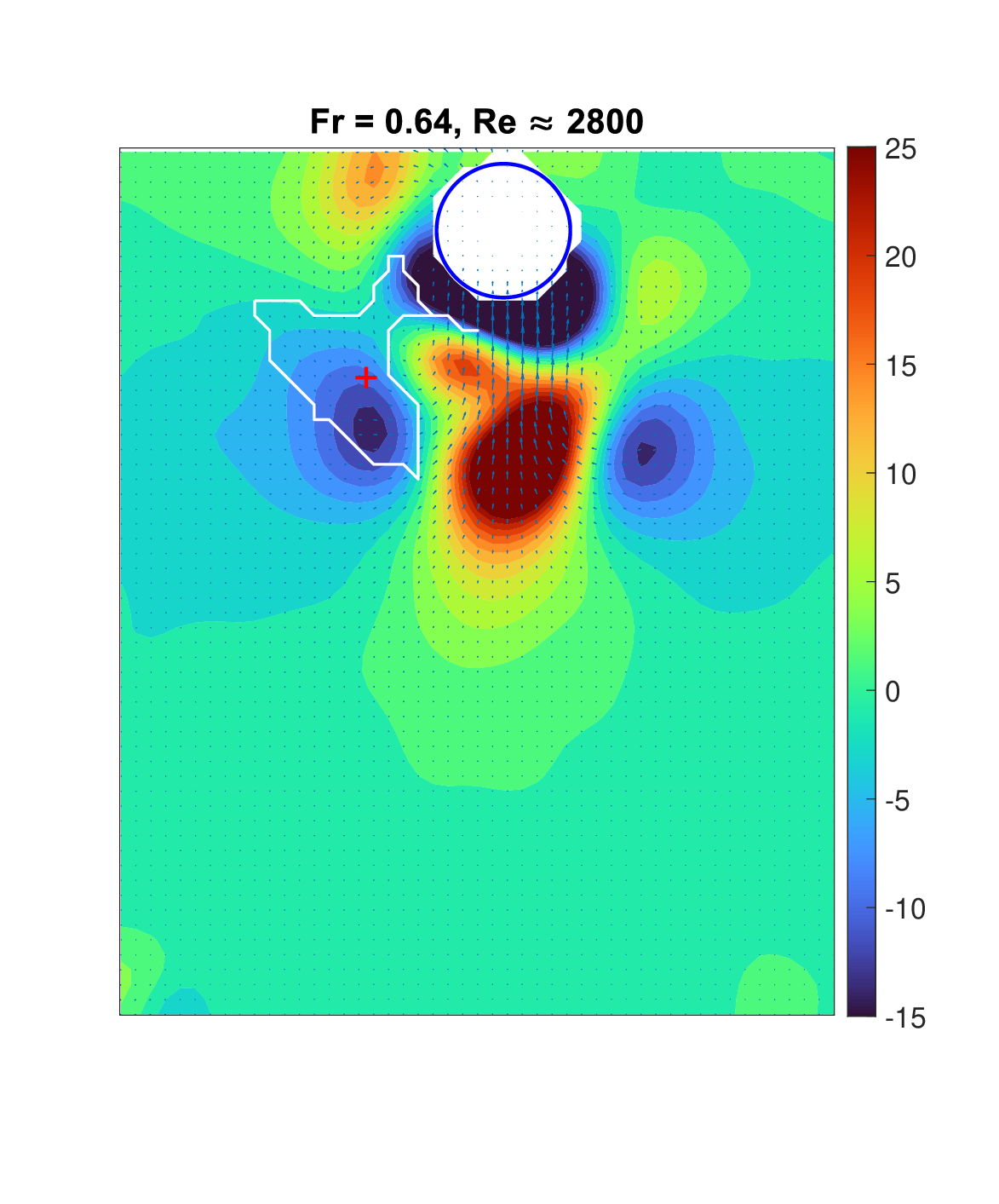}
    \caption{Identification of vortex at $y_{cyl}^* = 1$ for Fr =0.64, Re $\equiv 2800$.The white boundary shows the vortex-identified region. Vorticity field example. Darker shading corresponds to stronger vorticity magnitudes. The black spot beneath the cylinder is an imaging shadow and not a physical vortex.}
    \label{fig:vortex2}
\end{figure}

Section~\ref{sec:ResultsDiscussion}.
Representative masked regions for two Froude numbers appear in
Figures~\ref{fig:vortex1} and~\ref{fig:vortex2}.

\end{appendices}

\clearpage
\bibliography{sn-bibliography}

\end{document}